\documentclass[apj,twocolumn,chicago]{emulateapj}

\usepackage{amsmath}
\usepackage{float}
\usepackage{nicefrac}
\usepackage{rotating}
\usepackage{rotfloat}
\usepackage[none]{hyphenat}
\usepackage{enumitem}
\usepackage{afterpage}
\usepackage{datetime}
\usepackage{color}
\usepackage{url}
\usepackage[flushleft]{threeparttable}

\usepackage[colorlinks]{hyperref}
\usepackage{etoolbox}

\hypersetup{breaklinks=true,colorlinks=true,linkcolor=blue,citecolor=blue,filecolor=blue,urlcolor=blue}
\makeatletter
\patchcmd{\NAT@citex}
  {\@citea\NAT@hyper@{%
     \NAT@nmfmt{\NAT@nm}%
     \hyper@natlinkbreak{\NAT@aysep\NAT@spacechar}{\@citeb\@extra@b@citeb}%
     \NAT@date}}
  {\@citea\NAT@nmfmt{\NAT@nm}%
   \NAT@aysep\NAT@spacechar\NAT@hyper@{\NAT@date}}{}{}

\patchcmd{\NAT@citex}
  {\@citea\NAT@hyper@{%
     \NAT@nmfmt{\NAT@nm}%
     \hyper@natlinkbreak{\NAT@spacechar\NAT@@open\if*#1*\else#1\NAT@spacechar\fi}%
       {\@citeb\@extra@b@citeb}%
     \NAT@date}}
  {\@citea\NAT@nmfmt{\NAT@nm}%
   \NAT@spacechar\NAT@@open\if*#1*\else#1\NAT@spacechar\fi\NAT@hyper@{\NAT@date}}
  {}{}

\usepackage{array}
\newcolumntype{C}[1]{>{\centering\let\newline\\\arraybackslash\hspace{0pt}}m{#1}}

\def\aj{AJ}
\def\araa{ARA\&A}
\def\apj{ApJ}
\def\apjl{ApJ}
\def\apjs{ApJS}
\def\ao{Appl.~Opt.}
\def\apss{Ap\&SS}
\def\aap{A\&A}

\def\mnras{MNRAS}

\def\pasa{Publ.~Astron.~Soc.~Australia}
\def\pasp{PASP}

\def\rmxaa{RMxAA}

\def\arcdeg{\hbox{$^\circ$}}\def\arcmin{\hbox{$^\prime$}}
\def\arcsec{\hbox{$^{\prime\prime}$}}

\newcommand{\foiii}{[O\,{\sc iii}]}

\newcommand{\fsii}{[S\,{\sc ii}]}

\newcommand{\fnii}{[N\,{\sc ii}]}

\newcommand{\ovi}{O\,{\sc vi}}

\newcommand{\nevii}{Ne\,{\sc vii}}

\newcommand{\civ}{C\,{\sc iv}}

\newcommand{\heii}{He\,{\sc ii}}

\newcommand{\ha}{H$\alpha$}
\newcommand{\hb}{H$\beta$}

\newcommand{\bc}{\color{blue}}

\makeatletter
\patchcmd{\frontmatter@RRAP@format}{(}{}{}{}
\patchcmd{\frontmatter@RRAP@format}{)}{}{}{}
\renewcommand\Dated@name{}
\makeatother

\newcommand{\altaffilmarkc}[1]{\altaffilmark{\bc #1}}
\newcommand{\emailc}[1]{{\bc #1}}

\shorttitle{Inner Excitation Regions of NGC~5189}
\shortauthors{Danehkar et al.}

\begin{document}

\title{Mapping Excitation in the Inner Regions of the Planetary Nebula NGC~5189\\ Using \textit{HST} WFC3 Imaging}


\author{Ashkbiz~Danehkar\altaffilmarkc{1}, 
Margarita Karovska\altaffilmarkc{1}, 
W.~Peter Maksym\altaffilmarkc{1}, 
and Rodolfo Montez Jr.\altaffilmarkc{1}}
\affil{\altaffilmark{1}\,Smithsonian Astrophysical Observatory, 60 Garden Street, Cambridge, MA 02138, USA;
\emailc{ashkbiz.danehkar@cfa.harvard.edu}
}

\date[ ]{\footnotesize\textit{Received 2017 October 2; accepted 2017 November 28}}

\begin{abstract}
The planetary nebula (PN) NGC~5189 around a Wolf--Rayet [WO] central star demonstrates one of the most remarkable complex morphologies among PNe with many multi-scale structures, showing evidence of multiple outbursts from an AGB progenitor. In this study we use multi-wavelength  \textit{Hubble Space Telescope} Wide Field Camera 3 (WFC3) observations to study the morphology of the inner $0.3$\,pc\,$\times 0.2$\,pc region surrounding the central binary that appears to be a  relic of a more recent outburst of the progenitor AGB star. We applied diagnostic diagrams based on emission line ratios of H$\alpha$\,$\lambda$6563, \foiii\,$\lambda$5007, and \fsii\,$\lambda\lambda$6717,6731 images to identify the location and morphology of low-ionization structures  within the inner nebula. We distinguished two inner, low-ionization envelopes from the ionized gas, within a radius of 55 arcsec ($\sim 0.15$ pc) extending from the central star: a large envelope expanding toward the northeast, and its smaller counterpart envelope in the opposite direction toward the southwest of the nebula. These low-ionization envelopes are surrounded by a highly-ionized gaseous environment. We believe that these low-ionization expanding envelopes are a result of a powerful outburst from the  post-AGB star that created shocked wind regions as they propagate through the previously expelled material along a symmetric axis. Our diagnostic mapping using high-angular resolution line emission imaging can provide a novel approach to detection of low-ionization regions  in other PNe, especially those showing a complex multi-scale morphology.
\end{abstract}


\section{Introduction}
\label{ngc5189:introduction}

Planetary nebulae (PNe) are ionized hydrogen-rich shells, which are generated by strong mass-loss (superwinds) from low- to intermediate-mass progenitor star at the end of the asymptotic giant branch (AGB) phase. Ultraviolet radiation from hot degenerate cores fully ionizes the ejected shells during the post-AGB phase, and produces photo-ionized nebulae. Multiwavelength emission-line imaging observations illustrate how PNe appear in different excitation regimes \citep[see e.g.][]{Miranda1999,Sahai2007,Sahai2011}. As the degenerate core evolves from the AGB phase toward the white dwarf phase, it becomes hotter and its radiation pressure drives the stellar wind, leading to hydrodynamic shaping of the surrounding  nebular shell. Morphological studies of PNe provide important clues about the mass-loss process at the end of the AGB phase and the PN evolution during the post-AGB phase \citep[see e.g.][]{Stanghellini1993,Balick2002,Schoenberner2005a,Schoenberner2005b,Kwok2010,Steffen2013}.

NGC~5189 ($=$\,PN~G307.2$-$03.4\,$=$\,Hen~2-94\,$=$\,VV~65\,$=$ Sa 2-95) is a complex PN with multiple point-symmetric outflows or knots. The angular dimensions of NGC~5189 were measured to be about $163.4\arcsec \times 108.2\arcsec$ \citep[$0.43$\,pc $\times 0.29$ pc using $D=546$\,pc][]{Stanghellini2008} 
at the 10\% level of the \ha\ peak surface brightness \citep{Tylenda2003}. More than a half century ago, \citet{Evans1950} first described NGC~5189 as a planetary nebula or a massive nebula with remarkable knotted structures with no bright central star. \citet{Phillips1983} later suggested that it may contain multiple pairs of knots revealed in the narrow-band images of the \fnii, \ha, and \foiii\ emission lines. Moreover, Fabry-Perot imaging \fnii\ observations of NGC~5189 indicates that the nebula contains a dense ring expanding at 25 km\,s$^{-1}$ from the central star, with inclination of $78^{\circ}$ relative to the line of sight \citep{Reay1984}. \citet{Gonccalves2001} classified NGC~5189 as a bipolar PN with multiple pairs of knots inside the main structure with velocities similar to those of the surrounding gas.
More recently, \citet{Sabin2012} suggested that NGC~5189 is a quadrupolar PN containing a dense and cold infrared torus, and multiple point-symmetric structures.

\begin{table*}
\begin{center}
\caption[]{Hubble Observations of NGC~5189 taken on 2012 July 6 (Program
12812, PI: Z. Levay). \label{tab:obs:log}}
\begin{tabular}{ccccccl}
  \hline\hline\noalign{\smallskip}
Dataset  &	Exposure (s) & Instrument & Filter & $\lambda_{\rm peak}$(\AA) & $\Delta\lambda$(\AA) & Note	\\
\noalign{\smallskip}
\tableline
\noalign{\smallskip}
IBXL04010 & 4200 & WFC3/UVIS & F673N  & 6731 & 42 &  \fsii \\
\noalign{\smallskip}
IBXL04020 & 3900 & WFC3/UVIS & F657N  & 6573 & 41 &  H$\alpha$+\fnii \\
\noalign{\smallskip}
IBXL04040 & 8400 & WFC3/UVIS & F502N  & 5013 & 27 &  \foiii \\
\noalign{\smallskip}
IBXL04030 & 360  & WFC3/UVIS & F814W  & 8353 & 657 &  continuum \\
\noalign{\smallskip}
IBXL04050 & 300  & WFC3/UVIS & F606W  & 5956 & 663 &  continuum \\
\noalign{\smallskip}\hline
\end{tabular}
\end{center}
\end{table*}

The central star (CSPN) of NGC~5189 has been studied by a number of authors. It has been classified as 
a stellar type with broad \ovi\ and less broad \heii\ emission lines \citep{Blanco1968}, 
an ``\ovi\ sequence'' object \citep{Smith1969}, [WC\,2] spectral type \citep{Heap1982,Mendez1982}, [WO] \citep{Polcaro1997}, and [WO1] spectral type \citep{Crowther1998}. The \civ\ P-Cygni feature from the \textit{IUE} UV spectrum shows a terminal wind velocity of $-1540$ km\,s$^{-1}$ \citep{Feibelman1997}, while the line width from the optical spectrum suggests a maximum wind velocity of $-2800\pm200$ km\,s$^{-1}$ \citep{Polcaro1997}. Recently, \citet{Keller2014} derived a terminal wind velocity of $-2500\pm250$ km\,s$^{-1}$ and a stellar temperature of 165\,kK from from far-UV \nevii\ and \ovi\ P-Cygni profiles in the \textit{FUSE} spectrum of NGC 5189.

More recently, \citet{Manick2015} discovered significant periodic
variability which is associated with a binary having an orbital period of $4.04\pm 0.1$ days and a companion mass of $\geq 0.5$ M$_{\odot}$ or $0.84$ M$_{\odot}$ (at the orbital inclination of $40^{\circ}$). The complex morphology of NGC 5189 dominated by multiple, low-ionization structures (LISs) follows the trend of post common-envelope PNe outlined by \citet{Miszalski2009}. Interactions between the [WR] star of NGC~5189 and its companion may be responsible for its fast stellar [WR] winds \citep{Manick2015}.

In this work, we use high-resolution \textit{Hubble Space Telescope} images to analyze the spatially resolved inner structures of NGC\,5189  \citep[$120\arcsec\times90\arcsec$, or $0.32$\,pc $\times 0.24$ pc using a distance of $546$\,pc adopted from][]{Stanghellini2008} with a diagnostic classification for excitation regions. This paper is organized as follows. In Section \ref{ngc5189:observations}, we describe
the observations and data analysis. In Section \ref{ngc5189:diagnostic}, we present the results, including newly identified low-ionization envelopes within the nebula. 
In Section \ref{ngc5189:discussion}, we discuss our results, and in
Section~\ref{ngc5189:conclusions} we summarize the conclusions of this study.

\section{Observations and Data Analysis}
\label{ngc5189:observations}

The \textit{Hubble Space Telescope} (\textit{HST}) observations of NGC\,5189 were obtained on 2012 July 6 (Program 12812, PI: Z.~Levay) using the Wide Field Camera 3 (WFC3) instrument through the Ultraviolet-Visible (UVIS) channel configuration with the F673N, F657N, F502N, F814W, and F606W filters. These observations listed in Table~\ref{tab:obs:log} 
were taken with exposures of 4200, 3900, 8400, 360, and 300 sec, respectively. The WFC3 images were downloaded from the archive, and reprocessed using the {\tt AstroDrizzle} task of the \textit{HST} DrizzlePac v2.0 software package\footnote{\url{http://drizzlepac.stsci.edu/}} \citep{Gonzaga2012} in PyRAF v2.1.14, a Python-based command language for IRAF tasks. The sky background is automatically calculated for each image and subtracted using the function {\tt AstroDrizzle} at a sampling sky width of $0.1 \sigma$ and clipping limit of $1\sigma$. While the sky subtraction is performed, pixels within $1\sigma$ of the sky median value are excluded from the drizzled images. The WFC3 images were then aligned with the drizzled F814W image as a reference frame using the DrizzlePac functions {\tt TweakReg} and {\tt TweakBack}, and cosmic rays were removed from the images using the {\tt AstroDrizzle} task from the DrizzlePac tools. 

The systemic radial velocity of NGC 5189 is $-8\pm4$ km\,s$^{-1}$ \citep{Manick2015}, so the velocity shift is negligible, and the F657N, F502N, and F673N filters do cover the \ha\,$\lambda$6563, \foiii\,$\lambda$5007, and \fsii\,$\lambda\lambda$6716,6731 emission lines, respectively.

\begin{figure*}
\begin{center}
\includegraphics[width=3.5in]{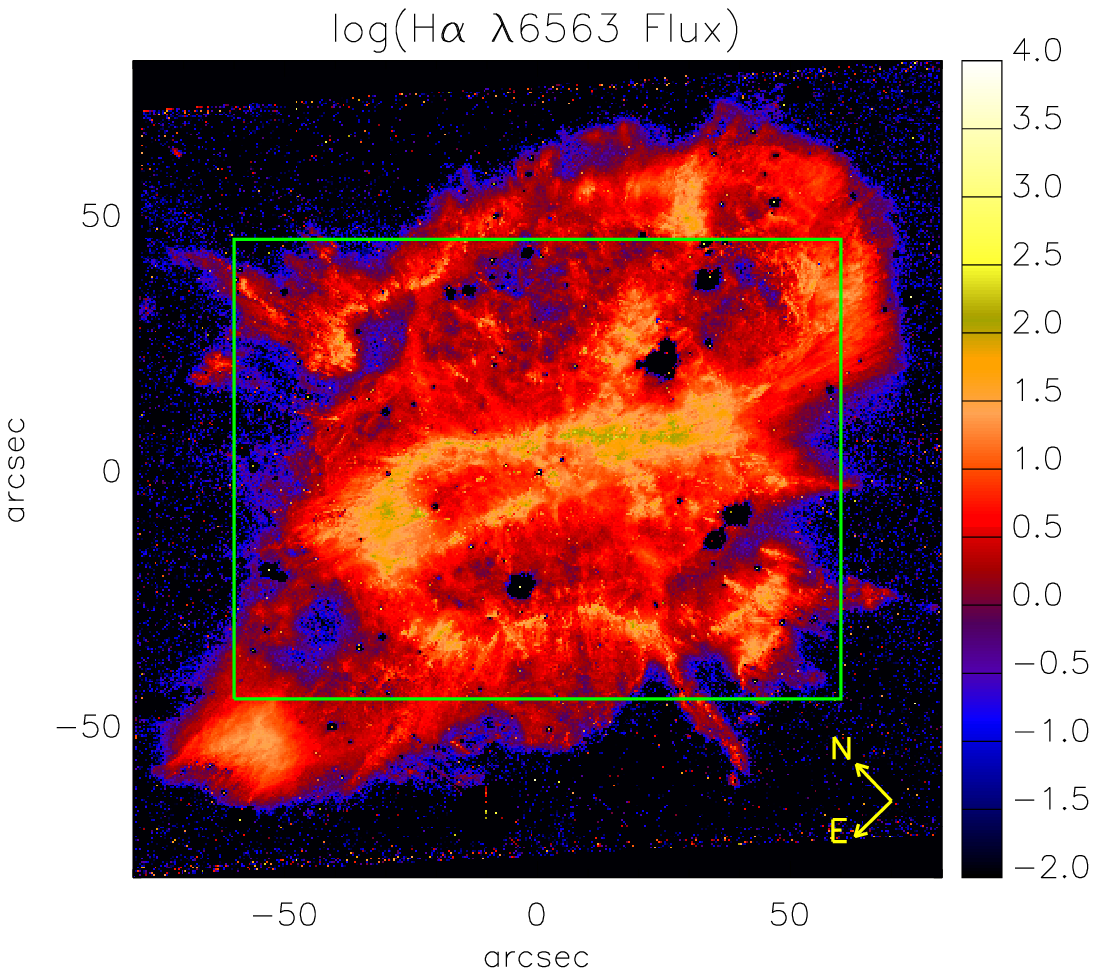}%
\includegraphics[width=3.5in]{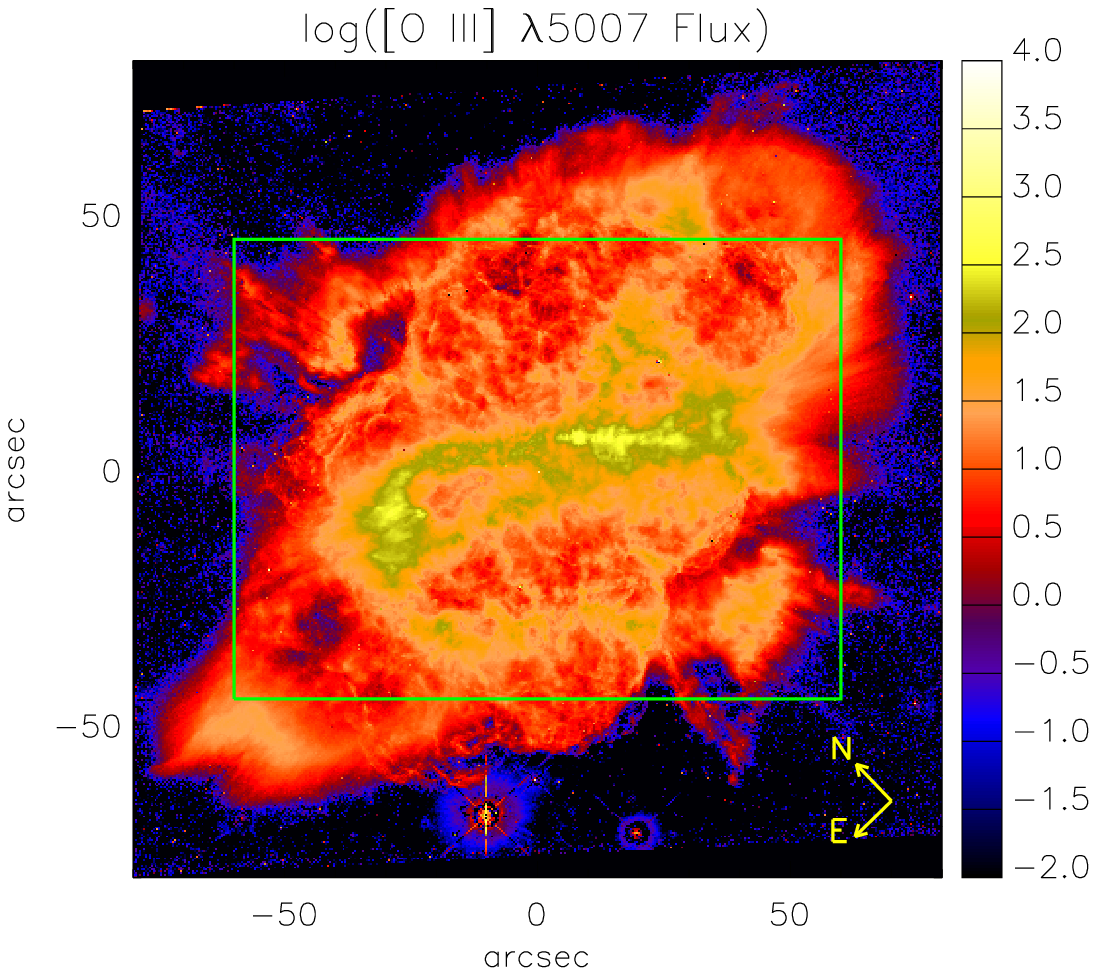}\\
\includegraphics[width=3.5in]{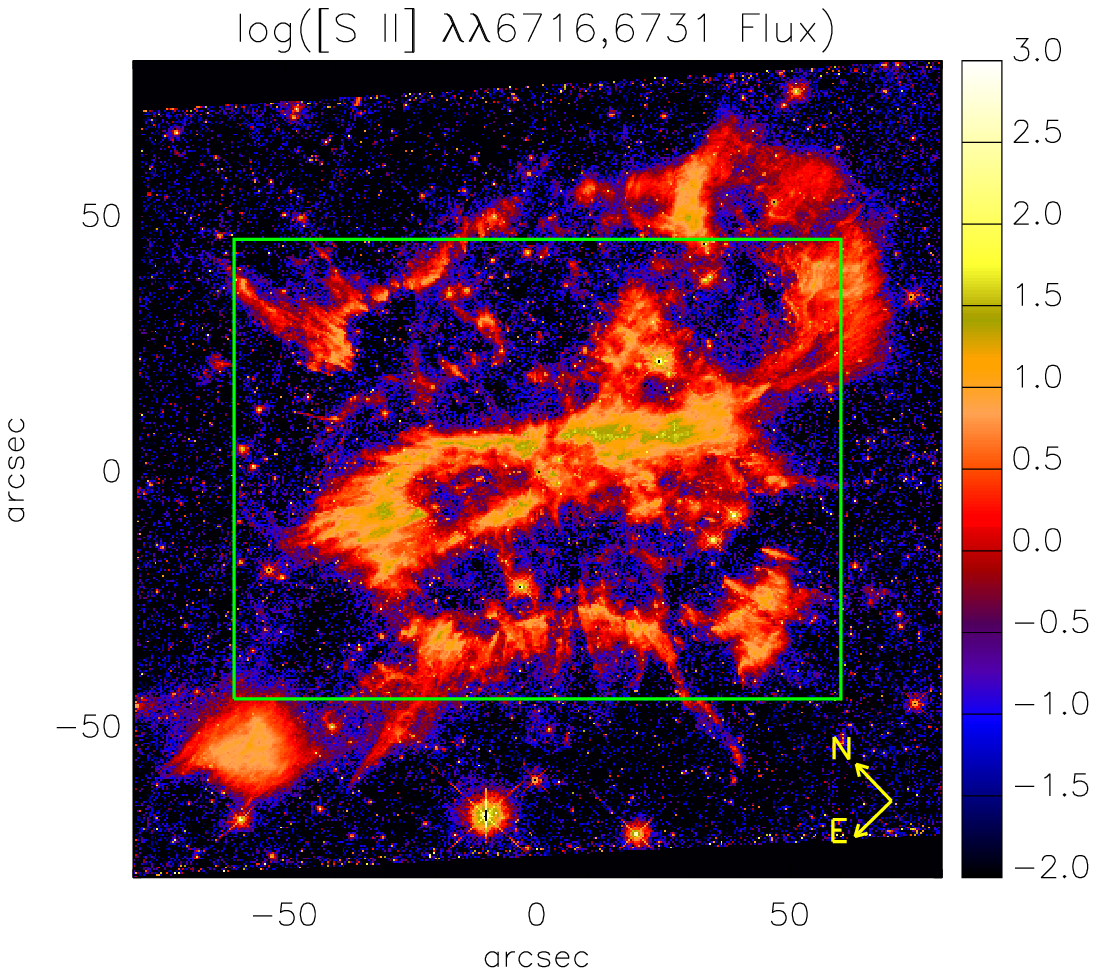}%
\includegraphics[width=3.5in]{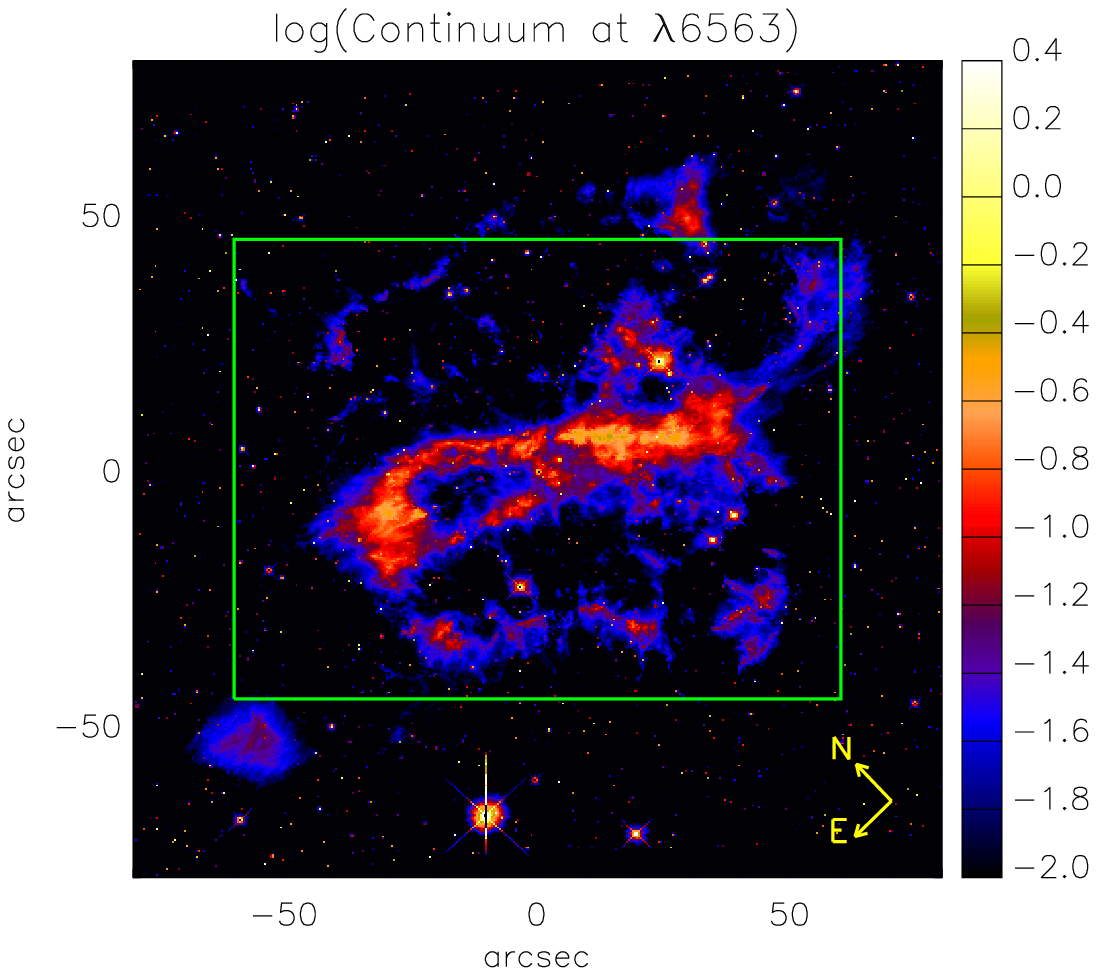}
\caption{From left to right, and top to bottom, dereddened, continuum-subtracted flux maps on
logarithmic scales (unit in $10^{-15}$~erg\,cm${}^{-2}$\,s${}^{-1}$\,arcsec${}^{-2}$) of \ha\,$\lambda$6563, \foiii\,$\lambda$5007, \fsii\,$\lambda\lambda$6716,6731 emission lines, and continuum flux-density on a logarithmic scale (unit in $10^{-15}$~erg\,cm${}^{-2}$\,s${}^{-1}$\,arcsec${}^{-2}$\,{\AA}${}^{-1}$) estimated at $\lambda$6563. 
Images include only pixels with at least statistically significant 1$\sigma$ sky.  The green rectangle indicates the $120\arcsec\times90\arcsec$ region of NGC~5189 used for detailed diagnostic mappings. Note that the upper value of the colorbar for the continuum map has been scaled down to show the faint nebular continuum. 
\label{fig:ngc5189:images}%
}
\end{center}
\end{figure*}

\subsection{Emission Line Mapping}

Emission line mapping from narrow-band imaging requires careful attention due to the potential contamination from undesired emission lines in a filter band-pass and the continuum emission. For NGC 5189, we used the available narrow and wide filters to remove these contaminants. Our process relied on the available narrow-filter observations (F673N, F657N, and F502N) to produce spatially-resolved flux-density maps of the \foiii\,$\lambda$5007{\AA}, \ha\,$\lambda$6563{\AA} and \fsii\,$\lambda\lambda$6716,6731{\AA} emission lines, and the wide-filter observations (F606W and F814W) for continuum subtraction. The images were flux-calibrated using the PHOTFLAM descriptor value, which converts count numbers (electrons sec$^{-1}$) into physical flux units of erg\,cm$^{-2}$\,s$^{-1}$\,{\AA}$^{-1}$\,arcsec$^{-2}$ (pixel size is $0.0396^{2}$ arcsec$^{2}$). 
The interpolated and extrapolated F606W and F814W flux-density maps were used to estimate the F657N and F502N continuum fluxes. These were then used to subtract from the F657N and F502N flux-density maps, while the stellar contaminated pixels were excluded from the F673N flux-density map using a flux limit according to the stellar mean flux.
The flux-density images of the narrow-filter bands were then transformed to mean flux maps by using the RMS bandwidths (PHOTBW header keyword).

The F657N filter bandpass also includes the \fnii\,$\lambda\lambda$6548,6584 emission lines, so the \ha\,$\lambda$6563 emission flux is a fraction of the flux maps derived from the F657N band. Since both the N$^{+}$ and S$^{+}$ ions have roughly similar ionization energies, the \fnii\ stratification layer should follow the \fsii\ morphology \citep[see e.g. Fig.~4 in][]{Danehkar2014}. The ratio \fnii/\fsii\  is measured to be about $ \approx 6$ from a $1 \arcsec \times 5 \arcsec$ slit \citep{Garcia-Rojas2012} that was taken from a bright knot with the MIKE
spectrograph (exposure times of 120 and 1800 sec and airmass of 1.25), so the slit covers a tiny fraction of the nebula. 
Our current analysis suggests that this value is too large for deriving an \fnii\ distribution from the F673N band, since the resulting \ha\ flux is inconsistent with the \foiii/\ha\ flux ratio of 4.05 reported by \citet{Garcia-Rojas2012} and 3.95 measured by \citet{Kingsburgh1994}. 
The \foiii $\lambda$5007/\ha\ surface brightness shows little variation across the nebula (see Fig.~\ref{ngc5189:oiii_ha:maps}), so a constraint on the \foiii/\ha\ flux ratio ($\approx 4$) can be employed to estimate a correct value of the \fnii/\fsii\ ratio. 
Through an iterative process we found that the ratio \fnii\ $=$ $4 \times $ \fsii\ yields a mean value of the dereddened \foiii\ $\lambda$5007 flux over the \fnii-corrected \ha\ flux of 3.98 for the $120\arcsec\times90\arcsec$ region (shown in Fig.~\ref{fig:ngc5189:images}), which agrees with both \citet{Kingsburgh1994} and \citet{Garcia-Rojas2012}. Hence, we adopted the \fnii/\fsii\ ratio of 4 to produce a map of the \fnii\ emission. We used this synthesized \fnii\ map to remove the \fnii\ contamination from the F657N band, leaving a pure \ha\  line-emission image.
Note that the \fnii/\fsii\ ratio will not always be constant when both shock-ionization and photo-ionization are present due to the shock-excitation dependence of this ratio. 
Additionally, the \textit{HST} WFC3 images of the Ring Nebula (NGC\,6720), calibrated using ground-based spectra showed that
the F673N filter has an uncertainty around 10 percent while the F502N and F658N fliters do not require any corrections \citep{ODell2013}. Unfortunately, the calibration corrections derived by \citet{ODell2013} require the F547M filter that could not be performed using the available filter set.

Following the method by \citet{Zeidler2015}, we estimated the continuum flux from the F606W and F814W fluxes. We estimated the continua of the \ha\ and \foiii\
emission lines by interpolating and extrapolating the fluxes of the F606W and F814W
images at the central wavelengths of the F657N and F502N bands, respectively.   
Although the F606W filter bandpass includes some nebular emission
lines such as \ha\ and \foiii, the short exposure time (300\,sec in F606W when compared to 3900\,sec in F657N and 8400\,sec in F502N) and the F606W-F814W interpolation \citep[also used by][]{Zeidler2015} prevent any large contributions from
these nebular lines. For example, see the faint nebular continuum in Figure \ref{fig:ngc5189:images}  (bottom-right panel). 
The estimated continuum of each image was then subtracted for a better removal of the stellar contamination from the narrow-band images in order to get the final, pure line-emission image. 
The removal worked well for the image in the brightness range between unsaturated and brighter than 1-$\sigma$ sky, but it has problems with saturated objects.
The continuum subtraction might fail for extremely faint emission ($< 1\sigma$ sky).  
As a result, the continuum reduction is not reliable for obtaining a pure line-emission \fsii\ image from the F673N, since the \fsii\ $\lambda\lambda$6716,6731 doublet is extremely weak. 
Instead, we used a flux limit for the F673N band which excludes those bright pixels associated with the stellar contamination. 
The flux ratio map \fsii/\ha\ is a key diagnostic in our analysis, but since those pixels associated with stellar contamination were already excluded by the continuum-subtraction in the \ha\ flux map, the ratio map is free of stellar contamination. 
The PHOTBW header keyword ($\Delta\lambda$ in Table~\ref{tab:obs:log}), which describes the RMS bandwidth (in {\AA}), was used to calculate the mean flux maps (in erg\,cm$^{-2}$\,s$^{-1}$\,arcsec$^{-2}$) from each calibrated, continuum-subtracted flux-density images (in erg\,cm${}^{-2}$\,s${}^{-1}$\,arcsec${}^{-2}$\,{\AA}${}^{-1}$). 

The continuum-subtracted flux maps were dereddened using the logarithmic extinction of $c({\rm H}\beta)=1.451 \times E(B-V)=0.47$ \citep{Garcia-Rojas2012}. To correct flux images for the interstellar extinction, we utilized the standard Galactic extinction law with $R_V \equiv A(V)/E(B-V)=3.1$ \citep{Seaton1979a,Howarth1983,Cardelli1989}. We also derive the H$\beta$ map from the H$\alpha$ map by adopting the reddened flux ratio $F({\rm H}\alpha)/F({\rm H}\beta)=4.187$ from \citet{Garcia-Rojas2012}. For the dereddening process, we assume that the extinction distribution is uniform, however, it could be inhomogeneous due to contributions from dust grains embedded inside the nebula. 

The final dereddened, continuum-subtracted flux maps are presented in Figure~\ref{fig:ngc5189:images}. These images are sky-subtracted, and pixels with values below 1$\sigma$ are masked 
to only include statistically significant pixels. 
Figure~\ref{fig:ngc5189:images} shows a pair of dense bright shells in the \ha\ and \foiii\ flux maps, which are  refereed as ``envelopes'' throughout this paper: one bright shell is extended from the central star toward the northeast of the nebula (${\rm PA} \approx 60^\circ$), while another smaller bright shell is extended from the central star toward the southwest of the nebula (${\rm PA} \approx 240^\circ$). The \foiii\ flux map almost looks similar to the \ha\ flux map, except for some small low emission regions in both sides of the dense bright envelopes in \ha\ emission. The \fsii\ flux map contains the same dense envelopes, however, the whole nebula is much fainter and contains several bright filamentary and knotty structures. The continuum density-flux presented in Figure~\ref{fig:ngc5189:images} (bottom-right panel) 
is the continuum contamination distribution estimated at the F657N central wavelength using interpolation between the F606W and F814W density-flux maps, and is used to subtract from the \ha\ flux map. A similar continuum density-flux was produced for \foiii\ as well. 

\section{Diagnostic Mapping Results}
\label{ngc5189:diagnostic}

An excitation diagnostic diagram consisting of \foiii/\ha\ versus \fsii/\ha\ was first produced by \citet{Phillips1999} to determine bow-shock regions in the bipolar outflows of the PN M2-9. Such an excitation diagnostic diagram was also used to distinguish between the shock- and photo-ionized regions in K\,4-47 \citep[][]{Gonccalves2004}, which is a PN consisting of a high-ionization core and a pair of LISs. To discriminate photoionized nebulae from shock-excited PNe, \cite{Raga2008} constructed a set of diagnostic diagrams, including \foiii/\ha\ versus \fsii/\ha, based on axisymmetric simulations of 
fast, dense LISs moving through a low-density environment, and away from an ionizing source. 
These diagnostic diagrams have been employed to distinguish low-ionization knots from photo-ionizated nebulae in numerous studies \cite[e.g.,][]{Gonccalves2009,Akras2016,Ali2017}. More recently, \citet{Akras2016} employed the \foiii/\ha\ versus \fsii/\ha\ excitation diagnostic diagram from \citet{Raga2008} to study a number of PNe with LISs.

\begin{figure*}
\begin{center}
\includegraphics[width=5in, trim = 0 0 0 0, clip, angle=0]{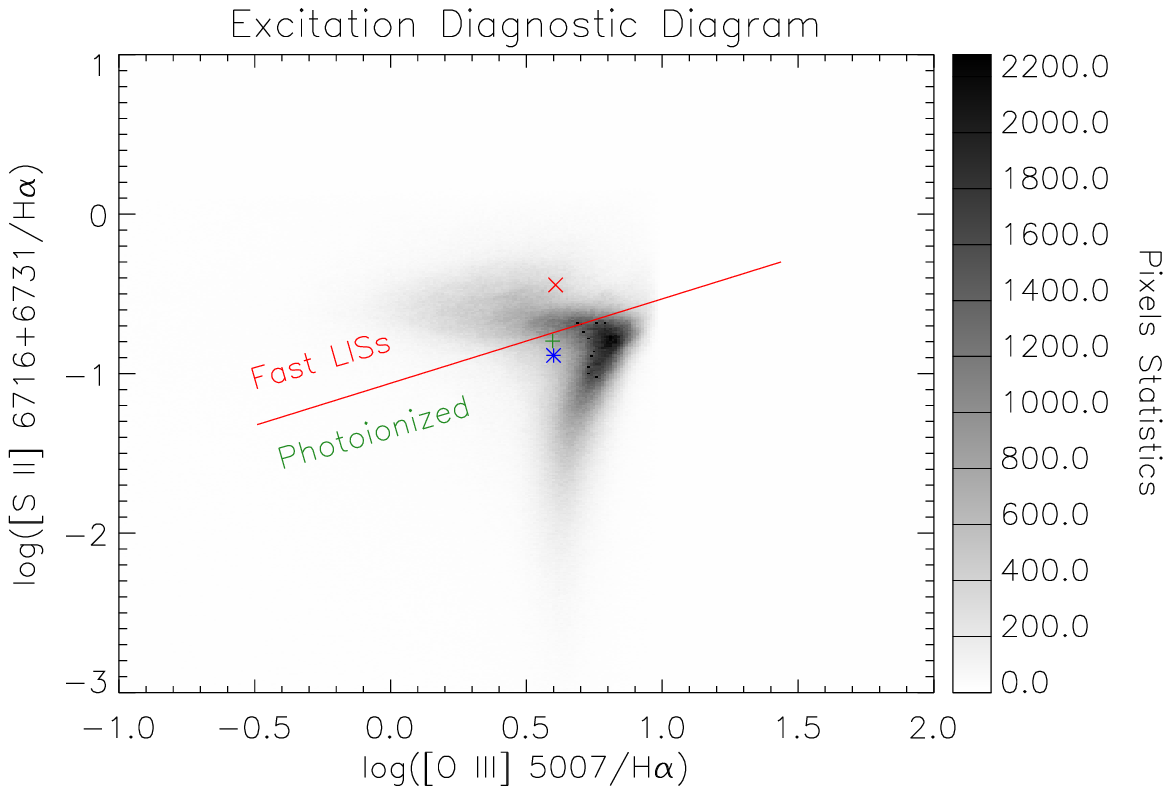}\\~\\
\includegraphics[width=5in, trim = 0 0 0 0, clip, angle=0]{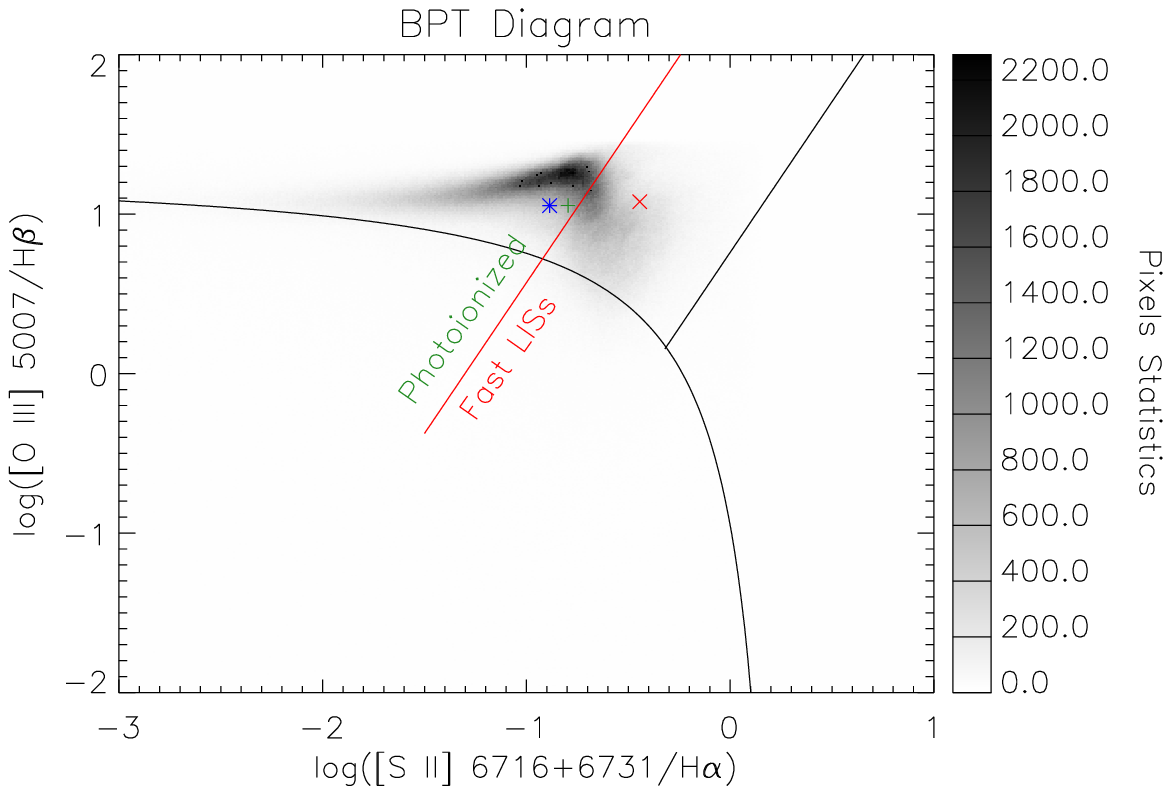}%
\caption{Excitation diagnostic diagram and BPT diagram of the inner region of NGC~5189, covering a $120\arcsec\times90\arcsec$ region (see Fig.~\ref{fig:ngc5189:images}). 
Top panel: Excitation diagnostic diagram presents logarithmic ratio maps of \foiii/\ha\ and \fsii/\ha. 
Bottom panel: BPT diagram presents logarithmic ratio maps of  \foiii/\hb\ and \fsii/\ha. Solid black lines show the boundaries of LINER-like and Seyfert-like activities from \citep{Kewley2006}. 
The solid red line depicts the nebular photon-shock dividing line in each panel chosen based on the shock models from \citet{Raga2008}, as described in the text. The star points ($\ast$) show the mean flux ratios of the $120\arcsec\times90\arcsec$ region, while the cross ($\times$) and plus ($+$) points depict the flux ratios from  \citet{Garcia-Rojas2012} and \citet{Kingsburgh1994}, respectively.
\label{fig:ngc5189:diagnostic}%
}
\end{center}
\end{figure*}

\begin{figure}
\begin{center}
\includegraphics[width=3.4in]{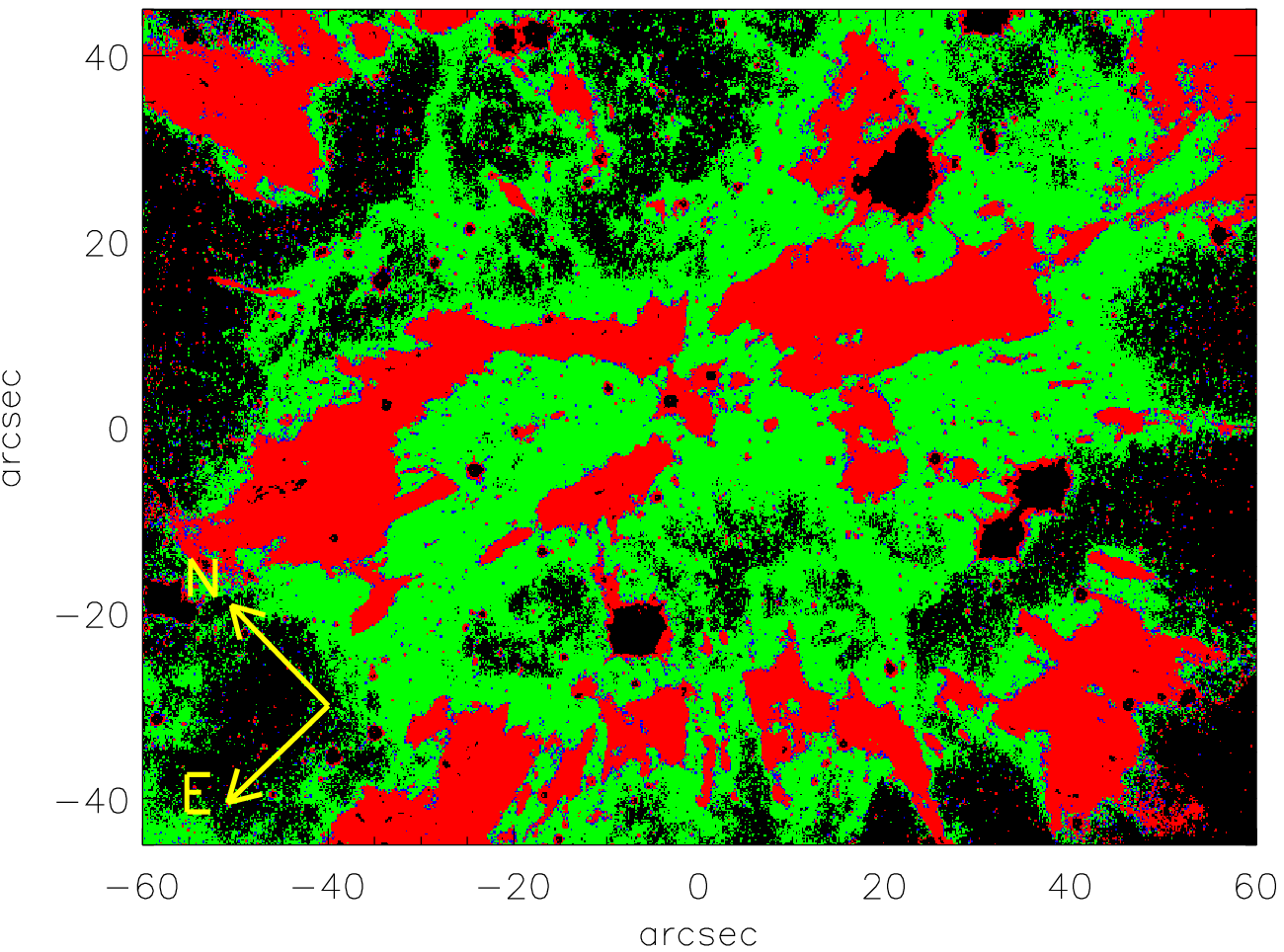}
\caption{Spatially-resolved diagnostic map of the inner region of NGC~5189, covering a $120\arcsec\times90\arcsec$ region centered on its [WO] central star, as shown in Fig.~\ref{fig:ngc5189:images}. The two pixel groups are color-coded according to their locations on the excitation diagnostic diagram in Fig.~\ref{fig:ngc5189:diagnostic}. Red pixels correspond to fast, low-ionization regions, and green pixels correspond to photo-ionized regions. Black pixels have either only one diagnostic line (\foiii\ or \fsii), or  one or both of diagnostic lines without at least $1\sigma$ of the mean value of the sky region. 
\label{fig:ngc5189:resolved}%
}
\end{center}
\end{figure}

Figure~\ref{fig:ngc5189:diagnostic} (top) presents an excitation diagnostic diagram with a 2-D histogram of log(\foiii/\ha) versus log(\fsii/\ha) plotted for $1\sigma$-masked WFC3 pixels extracted from the $120\arcsec\times90\arcsec$ region, centered on the [WO] central star of NGC~5189 (RA\,DEC/J2000: 13h33m32.9s $-$65$^{\circ}$58$\arcmin$27.1$\arcsec$).
We selected a $120\arcsec\times90\arcsec$ extraction region, oriented with a position angle of $\sim 45^{\circ}$ (from the north toward the east in the equatorial coordinate system) to focus on the filamentary structures around the central star. 
The spatial resolution of each WFC3 pixel is $0.0396\arcsec\times0.0396\arcsec$, corresponds to $\sim 6.75 \times 10^6$ pixels in the extracted region, but not all pixels are statistically $1\sigma$ significant and have both \fsii/\ha\ and \foiii/\ha\ flux ratios. 
To disentangle fast LISs from photoionized regions, we adopted two different regions of the excitation diagnostic diagram guided by the calculations presented in \cite{Raga2008}. Specifically, we delineated fast LISs and photoionized regions. This diagram is similar to the Baldwin--Phillips--Terlevich (BPT) diagram \citep{Baldwin1981}, which is used to distinguish Seyfert-type and low ionization emission line region (LINER) classifications of starbursts and active galactic nuclei (AGN) galaxies \citep{Kewley2001,Kewley2006}. Recently, the BPT diagram has been used to spatially resolve Seyfert-type and LINER-type activities of the inner region of the extended narrowline region (ENLR) of NGC 3393 \citep{Maksym2016,Maksym2017}. For comparison, in Figure~\ref{fig:ngc5189:diagnostic} (bottom) we also show the corresponding BPT diagram of NGC~5189, including the shock-excited and photo-ionized regions according to the excitation classification from \citet{Raga2008}.

We used the excitation diagnostic diagrams presented in Figure~\ref{fig:ngc5189:diagnostic} to delineate the fast, dense LISs from the photo-ionized medium of the inner region of the nebula NGC~5189.  
All the valid pixels, which possess both \fsii/\ha\ and \foiii/\ha\ flux ratios, are included in the excitation diagnostic diagrams.
We adopted a nebular photon-shock dividing line according to shock models \citep{Raga2008}. Although this division was not clearly defined by \citet{Raga2008}, we adopted a photon-shock dividing line that is parallel with the Seyfert--LINER classification line, $1.89 \log($\fsii/\ha$)+0.76 = \log($\foiii/\hb$)$, defined by \citet{Kewley2006}. We use this line as an empirical division between photo-ionization and potential shock-ionization regimes within the nebula, hereafter referred to as the `nebular photon-shock dividing line'. Our empirical estimate based on the shock models for the location of this dividing line is $1.89 \log($\fsii/\ha$)+2.46 = \log($\foiii/\hb$)$ for the BPT diagram (See Figure~\ref{fig:ngc5189:diagnostic}). 
For the excitation diagnostic diagram shown in Figure~\ref{fig:ngc5189:diagnostic} (top), the same dividing line corresponds to $1.89 \log($\fsii/\ha$)+2.0 = \log($\foiii/\ha$)$. 

Figure~\ref{fig:ngc5189:resolved} shows the results of classifying pixels based on the diagnostic map of NGC\,5189. Red indicates fast, low-ionization activity, and green
is typical of photo-ionized regions. We excluded pixels without at least $\sim 1\sigma$ of the mean value of the sky region, and without both the \fsii/\ha\ and \foiii/\ha\ diagnostic ratios.

\begin{figure*}
\begin{center}
\includegraphics[width=7.0in]{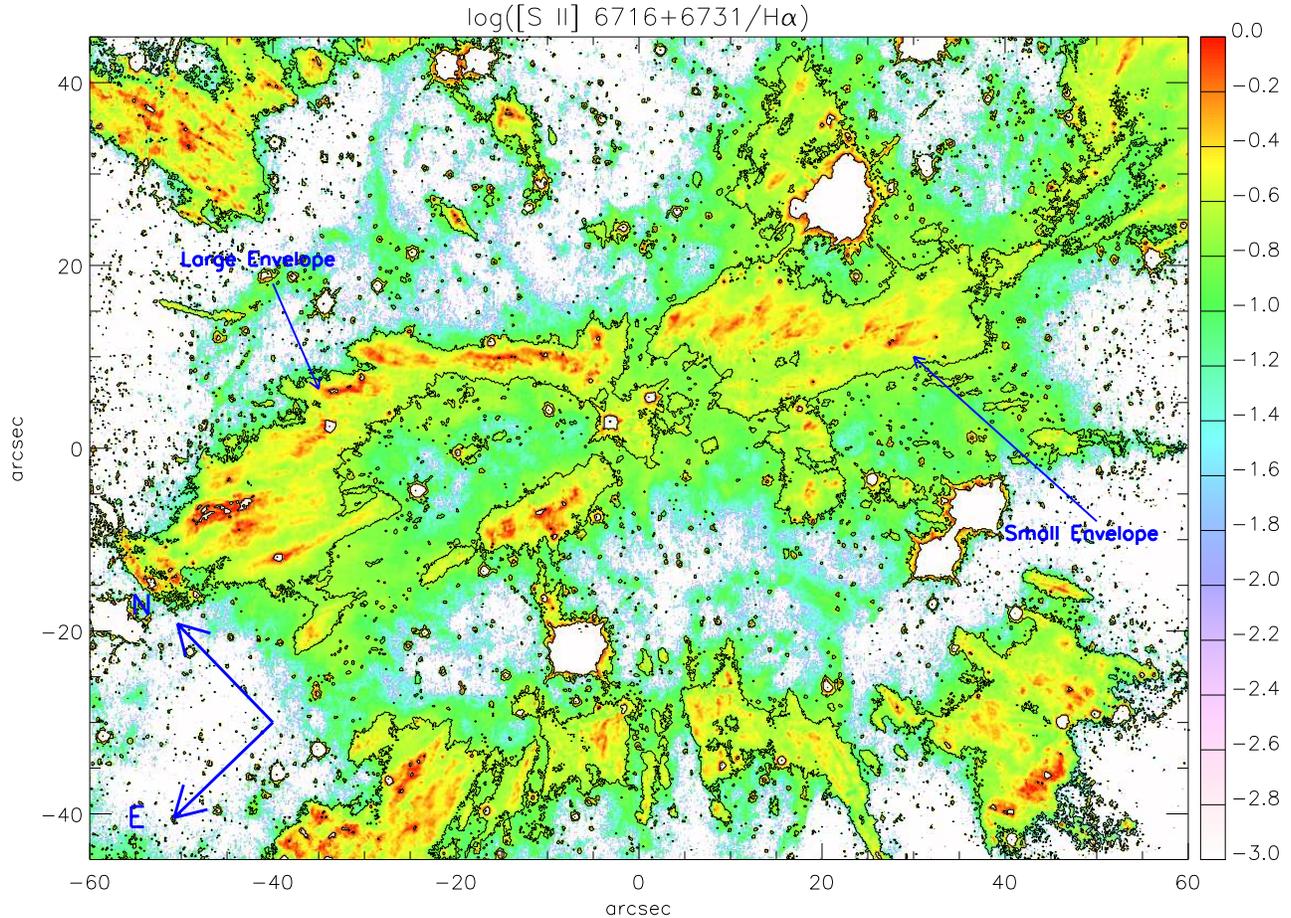}%
\caption{Logarithmic flux ratio map of \fsii/\ha\ produced from the \textit{HST} observations for the $120\arcsec\times90\arcsec$ region shown in Fig.~\ref{fig:ngc5189:images}. 
The contour lines illustrate the boundaries of photo-ionized region and fast, low-ionization region based on  the pixel classification seen in Fig.~\ref{fig:ngc5189:resolved}. 
The main morphological features, large low-ionization envelope, and its smaller counterpart are labeled.
}%
\label{ngc5189:sii_ha:maps}%
\end{center}
\end{figure*}

Based on the distributions in Figure~\ref{fig:ngc5189:diagnostic}, mean flux ratios measured from the $120\arcsec\times90\arcsec$ extracted region would be 
\fsii/\ha~=~$0.13 \pm 0.11$, \foiii/\ha~=~$3.98 \pm 1.63$ and \foiii/\hb~=~$11.32 \pm 4.62$ (indicated by $\ast$ in Fig.~\ref{fig:ngc5189:diagnostic}; errors correspond to the average absolute deviations). Flux ratios from the literature are \fsii/\ha~=~0.36, \foiii/\ha~=~4.05 and \foiii/\hb~=~11.97 \citep[indicated by $\times$ in Fig.~\ref{fig:ngc5189:diagnostic};][]{Garcia-Rojas2012}, 
\fsii/\ha~=~0.16, \foiii/\ha~=~3.95 and \foiii/\hb~=~11.34 \citep[indicated by $+$ in Fig.~\ref{fig:ngc5189:diagnostic};][]{Kingsburgh1994}. 
We note that \citet{Garcia-Rojas2012} used a $1\arcsec\times5\arcsec$ slit, whereas \citet{Kingsburgh1994} employed different slits with a total length of $18.4\arcsec$ and widths of $1\arcsec$ and $6.7\arcsec$. 
Despite the different slit configurations, the \foiii/\ha\ flux ratios are approximately the same. Our mean \fsii/\ha\ flux ratio of the $120\arcsec\times90\arcsec$ region is roughly similar to what is reported by \citet{Kingsburgh1994}, but a factor of $\sim 3$ lower than the flux ratio derived by \citet{Garcia-Rojas2012}. This discrepancy could be due to the shorter slit used by \citet{Garcia-Rojas2012}, and it is possible that the slit was placed on one of the nebular envelopes (see Figure~\ref{ngc5189:sii_ha:maps}) which is dominated by the fast, low-ionization regime. The longer and wider slits  ($18.4\arcsec\times1\arcsec$ and $18.4\arcsec\times6.7\arcsec$) employed by \citet{Kingsburgh1994} presented
the flux ratios close to our results.

\begin{figure*}
\begin{center}
\includegraphics[width=7.0in]{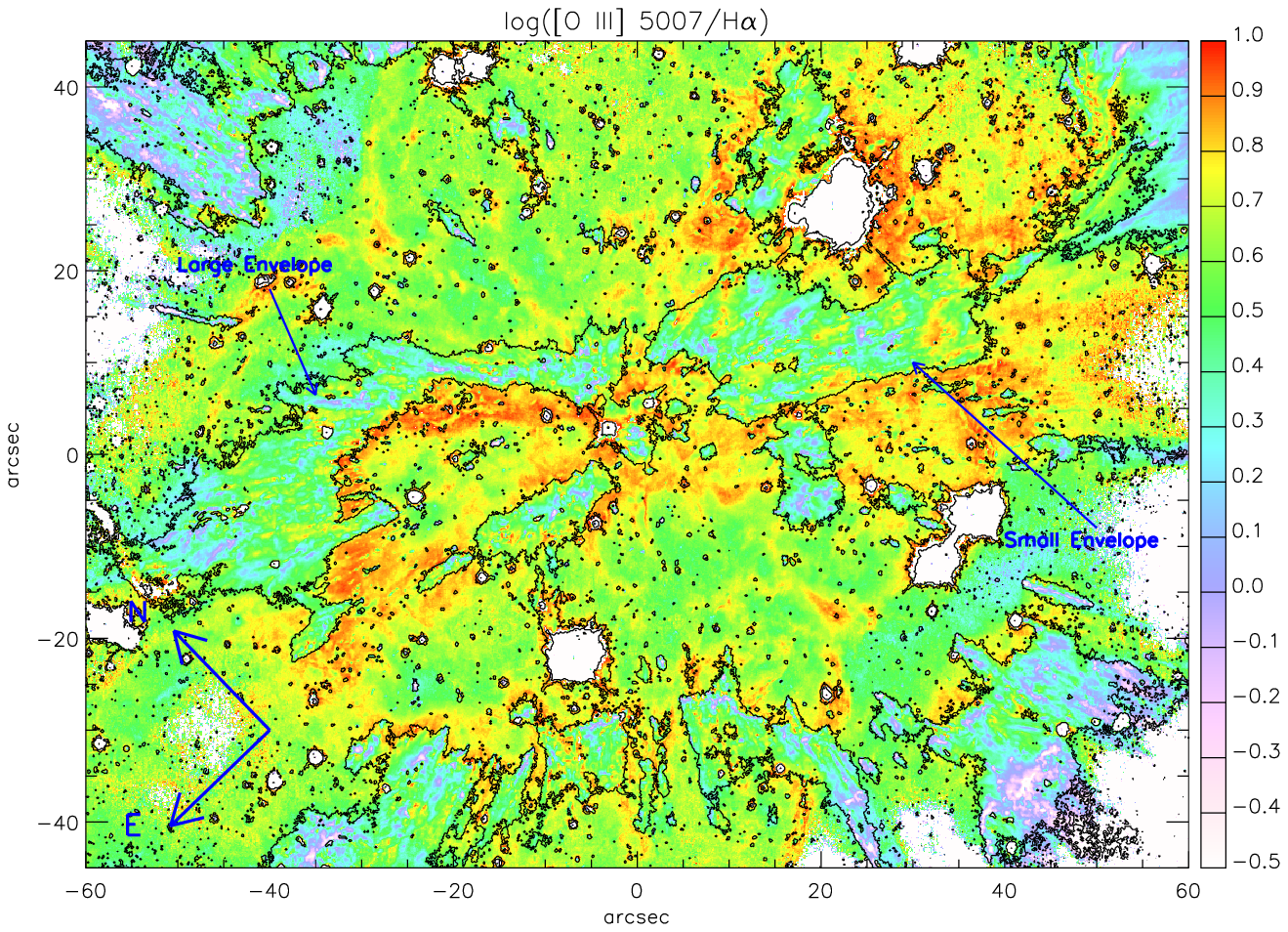}%
\caption{Logarithmic flux ratio map of \foiii/\ha\ produced from the \textit{HST} observations for the $120\arcsec\times90\arcsec$ region shown in Fig.~\ref{fig:ngc5189:images}. 
The contour lines illustrate the boundaries of photo-ionized region and fast, low-ionization region based on   the pixel classification seen in Fig.~\ref{fig:ngc5189:resolved}. 
The main morphological features, large low-ionization envelope, and its smaller counterpart are labeled.
}%
\label{ngc5189:oiii_ha:maps}%
\end{center}
\end{figure*}

The \fsii/\ha\ and \foiii/\ha\ logarithmic flux ratio maps are presented in Figures~\ref{ngc5189:sii_ha:maps} and \ref{ngc5189:oiii_ha:maps}, respectively, which were produced from the dereddened, continuum-subtracted flux maps shown in Figure~\ref{fig:ngc5189:images}. The boundaries of the photo-ionized and the fast, low-ionization regions based on Figure~\ref{fig:ngc5189:resolved} are illustrated as contour lines in both the flux ratio figures. The shock criterion  \citep[$\log($\fsii/\ha$)\gtrsim -0.4$;][]{Mathewson1973,Fesen1985} is satisfied in several places within the fast, low-ionization region, suggesting that the dense, filamentary structures are interacting with the surrounding low-density medium.

We can easily identify the following main morphological features of the inner excitation regions: 

(1) A large dense, low-ionization envelope with a maximum diameter of $\sim 55$ arcsec extended from the central star wherein the structures are expanding toward the northeast \citep[see e.g. long-slit observation;][]{Sabin2012}. 

(2) A smaller compact, dense low-ionization envelope with a maximum radius of $\sim 40 \arcsec$ is extended from the central star and is expanding toward the southwest, likely the large low-ionization envelope's counterpart.  

(3) These two low-ionization envelopes are surrounded by the highly ionized, low-density gas (showing $\log($\foiii/\ha$)\geqslant 0.8$ in Figure~\ref{ngc5189:oiii_ha:maps}).

(4) Multiple low-ionization filamentary and knotty structures in both of the two low-ionization envelopes.

As  only one long-slit kinematic observation is available for this region \citep[see][]{Sabin2012}, we could not constrain the 3-D geometry of these low-ionization envelopes through morphokinematic modeling. Additional long-slit observations are necessary in order to disentangle their 3-D morphological structures.

\section{Discussion}
\label{ngc5189:discussion}

Diagnostic mapping of the flux ratios \foiii/\ha\ and \fsii/\ha\ have allowed us to spatially resolve the fast, high-density LISs within the low-density photo-ionized environment of NGC~5189 (described in Section~\ref{ngc5189:diagnostic}). The regions of fast LISs is shown in Figure~\ref{fig:ngc5189:resolved}. There are two main LISs in the inner region of NGC\,5189 within $50\arcsec \times 50\arcsec$ ($0.13$\,pc $\times 0.13$\,pc at $D=546$\,pc) from the central star: the larger LIS expands toward the northeast of the nebula (${\rm PA} \approx 60^\circ$), while the smaller LIS expands toward the southwest (${\rm PA} \approx 240^\circ$). 

We notice that these inner LISs in NGC~5189 are also bright in both \ha\ and \foiii, in addition to \fsii\ (see Figure~\ref{fig:ngc5189:images}), contrary to the typical definition of LISs \citep[][and references therein]{Corradi1996,Gonccalves2001,Gonccalves2009}. These inner LISs are brighter in the \fsii/\ha\ flux ratio, and fainter in the \foiii/\ha\ flux ratio. However, these LISs are not fainter than the surrounding main nebula in the \fsii\ or \foiii\ emission absolute fluxes. 

The kinematic data of a long-slit passing through the central region covering the low-ionization envelopes indicate that they have maximum projected velocities up to 35--45 km\,s$^{-1}$ \citep[the slit b in][]{Sabin2012}.
According to these projected velocities, the inner LISs in NGC~5189 are not moving significantly faster than other regions, so their features may not be typical of the so-called fast, low-ionization emission regions (FLIERs) seen in around 50\% of PNe \citep[e.g.][]{Balick1993,Balick1994,Balick1998,Hajian1997,Perinotto2004,Danehkar2016}. Typically FLIERs appear  point-symmetric and their low-ionization outflows move supersonically with respect to the main nebula  \citep[][]{Balick1993,Balick1994}. Nevertheless, fast bipolar outflows detected in some PNe move faster than the main nebula, but their excitation characteristics may not correspond to LISs \citep[e.g.][]{Trammell1996,Corradi1997,Guerrero1998,Guerrero2008,Miranda2012,Danehkar2015,Fang2015}.

From Figure~\ref{fig:ngc5189:diagnostic}, the characteristics of these LISs within the nebula are typical of the shock-ionization, so their unprojected expansion velocity should be higher than the surrounding high-excitation material. These low-ionization envelopes expanding along an apparently symmetric axis may be caused by the past powerful outburst from the progenitor post-AGB star, plowing into the previously ejected material. 
While these envelopes are ionized by UV radiation from the hot central star  \citep[$T_{\rm eff}=165$~kK;][]{Keller2014}, their propagation through and interaction with the previously expelled matter makes the shocked wind regions that produce additional thermal energy for ionization \citep[see e.g.][]{Guerrero2013,Freeman2014,Dopita2017}. Studies of WFC3 images showed that flux ratio such as \foiii/\ha\ could be enhanced by bow-shock features \citep{Guerrero2013}. Moreover, X-ray \textit{Chandra} imaging observations suggested the presence of wind-shock-heated bubbles within PNe \citep{Freeman2014}.
Shock-ionization modeling demonstrates how a shock propagating at $\sim 40$ km/s into the pre-existing material can heat up them \citep{Dopita2017}, while there is also evidence for the shock-excitation of LISs in some PNe \citep{Ali2017}. Therefore, shock-ionization, in addition to photo-ionization, provides thermal energy that contributes to a deviation from the photoionization pattern in the diagnostic diagrams (on one side of the photon-shock dividing line in Fig.~\ref{fig:ngc5189:diagnostic}).

As Figure~\ref{ngc5189:sii_ha:maps} shows, the low-ionization envelopes contain several
filaments and knots that are bright in \fsii\ compared to H$\alpha$ and
\foiii. Numerical simulations of radiative shock models reveal that
radiative shock can form knots and filaments in a non-accelerated medium
such as PNe \citep{Walder1998a,Walder1998b}. It is possible that the knots
seen in the low-ionization envelopes of NGC~5189 provide the seeds for
cometary-like knots such as those seen in the Helix nebula \citep{ODell2004,Matsuura2009,Meaburn2013}. 
The kinematics and
composition of these early structures can provide valuable constrains on
the origin and evolution of knotty structure in PNe \citep{Redman2003}.

Currently, it is not fully understood how fast LISs and bipolar outflows are formed in PNe. It has been proposed that rotating stellar winds and strong toroidal magnetic fields generate equatorial density outflows \citep[e.g.,][]{Garcia-Segura1997,Garcia-Segura1999,Garcia-Segura2000,Frank2004}. Alternatively, axisymmetric superwind mass-loss could result from a common-envelope phase for a binary system consisting of a white dwarf or a low-mass companion \citep[e.g.,][]{Soker1994,Soker2006,Nordhaus2006,Nordhaus2007}. Previously, \citet{Miszalski2009} associated  complex morphologies, such as those seen in NGC 5189, with post common-envelope nebulae. Recently, the periodic variability of the central star of NGC~5189 was discovered and found to be related to binarity with a 4-day orbital period \citep{Manick2015}. Additionally, \citet{Bear2017} listed NGC~5189 among PNe with potential triple progenitors based on its complex morphology.
The low-ionization envelopes of NGC 5189 could therefore be the result of a binary or triple stellar evolutionary path. 

We note that as seen in Figure~\ref{fig:ngc5189:diagnostic} (bottom), NGC\,5189 seems to demonstrate  patterns of Seyfert-like activity \citep[see e.g.][]{Kewley2001,Kewley2006}. Interestingly, recent studies of early-type galaxies indicate that excitation classifications based on the BPT diagrams for a considerable fraction of them can be attributed to post-asymptotic giant branch (post-AGB) nuclei of planetary nebulae \citep{Annibali2010,Sarzi2010}. The ionization contribution from post-AGB central stars of planetary nebulae can therefore be partially responsible for LINER-like and Seyfert-like line ratios in galaxies. 
Although the diagnostics observed in this very high-resolution view of NGC\,5189 are typically Seyfert-like, the central stars of planetary nebulae, which are responsible for ionizing the nebulae, are typically orders of magnitude weaker than even the lowest luminosity AGN associated with LINERs \citep[see e.g.][]{Ho2008}. We therefore expect the local attenuation of the post-AGB Lyman continuum, e.g. by dust or geometric effects, to become very important on larger scales. A more complete and extended system should therefore be consistent with LINER-like ratios in galaxies, similarly to what is seen in attenuated AGN emission \citep[e.g.][]{Singh2013}.

\section{Summary and Conclusions}
\label{ngc5189:conclusions}

In this work, we used deep \textit{HST} WFC3 imaging of NGC\,5189 in \fsii, \foiii, and \ha\ to map low- and high-excitation regions within the inner ($120\arcsec\times90\arcsec$ or 0.32 pc $\times$ 0.24 pc) region of the nebula centered on its [WO] central star. The \textit{HST} images (see Figure~\ref{fig:ngc5189:images}) illustrate that NGC\,5189 contains multiple filamentary structures and several knots distributed inside the nebula. The inner region close to the central star includes filamentary loops, which are bright in the \ha, \foiii, and \fsii\ emission lines.

We employed diagnostic diagrams consisting of \foiii/\ha\ and \fsii/\ha\ ratios  to distinguish between fast, dense LISs (low-ionization structures) and low-density photoionized medium. For the excitation diagnostic  (\foiii/\ha\ vs.~\fsii/\ha\ ratios), the dividing line between the shock- and photo-ionized regions was adopted based on shock model simulations by \cite{Raga2008}, which demonstrates the shock-ionization effects of fast, dense material passing through a low-density environment. The adopted photon-shock dividing line of $1.89 \log($\fsii/\ha$)+2.0 = \log($\foiii/\ha$)$ is parallel with the Seyfert--LINER classification line \citep{Kewley2006} distinguishing between Seyfert-type and LINER-type activities in galaxies. We have used this approach to map fast LISs within the NGC\,5189 nebula. We identified two low-ionization envelopes in the inner $50\arcsec \times 50\arcsec$ ($0.13$\,pc $\times 0.13$\,pc) region of NGC\,5189 from the central star: 
one of them is a large envelope expanding toward the northeast (${\rm PA} \approx 60^\circ$), whereas its counterpart is a smaller envelope expanding toward the southwest of the nebula (${\rm PA} \approx 240^\circ$). 

Although the hot central star of NGC\,5189 provides UV radiation to ionize these low-ionization envelopes, propagation of dense LISs through the previously ejected materiel create wind-shock-heated features that 
could add a deviation from the photoionization pattern (see Fig.~\ref{fig:ngc5189:diagnostic}). 
We also noticed that the \textit{HST} diagnostic view of NGC 5189 corresponds to Seyfert-like activity in the BPT diagram (see Fig.~\ref{fig:ngc5189:diagnostic} bottom). Nebulae locally ionized by post-AGB stars can considerably contribute to LINER-like and Seyfert-like patterns in the BPT diagram on large scales of galaxies \citep[see e.g.][]{Annibali2010,Sarzi2010}.

Currently, only one long-slit spectrum of the central region is available, so 3-D kinematic structures of the LIS envelopes cannot properly be determined. 
Further high resolution kinematic observations, such as long-slit high-resolution spectroscopy at several positions and orientations in the $120\arcsec\times90\arcsec$ region will present further constraints to determine their 3-D morphological and kinematic characteristics. We also note that accurate reddening correction of \textit{HST} images for the interstellar extinction can be done if H$\beta$ $\lambda$4861 and radio maps are available. Further high-resolution deep \textit{HST} multiwavelength emission-line imaging and multiple long-slit spectroscopy will certainly provide crucial details about morpho-kinematic structures of inner, low-ionization envelopes in this planetary nebula.

\acknowledgments

We are grateful to Mr.~Zoltan~Levay and the Hubble Heritage Team who have obtained the NGC5189 \textit{HST} data. 
Based on observations made with the NASA/ESA \textit{Hubble Space Telescope}, obtained from the Data Archive at the Space Telescope Science Institute, which is operated by the Association of Universities for Research in Astronomy, Inc., under NASA contract NAS 5-26555. We thank the anonymous referee for constructive comments and corrections.

\textit{Facilities}: \facility{\textit{HST} (WFC3)}. 


\end{document}